\documentclass[referee,sn-mathphys]{sn-jnl}

\usepackage[version=4]{mhchem}
\usepackage{gensymb}
\usepackage{multibib}
\usepackage{miller}
\usepackage{overpic}
\usepackage{siunitx}
\usepackage{helvet}

\let\oldput\put
\def\put(#1,#2)#3{%
  \oldput(#1,#2){\fontfamily{phv}\selectfont #3}
}

\newcommand{\uvec}[1]{\boldsymbol{\hat{\textbf{#1}}}}
\definecolor{myorag}{RGB}{202, 86, 44}
\definecolor{myblue}{RGB}{41, 128, 128}
\definecolor{palat}{RGB}{142, 58, 89}
\definecolor{white}{RGB}{255, 255, 255}

\jyear{2022}%

\theoremstyle{thmstyleone}%
\theoremstyle{thmstyletwo}%

\theoremstyle{thmstylethree}%

\begin{document}

\title[Electric Field Controlled Mechanism for the Deflection of Skyrmions]{Electric Field Controlled Mechanism for the Deflection of Skyrmions}


\author*[1]{\fnm{Samuel H.} \sur{Moody}}\email{samuel.h.moody@durham.ac.uk}

\author[1]{\fnm{Matthew T.} \sur{Littlehales}}\email{matthew.t.littlehales@durham.ac.uk}

\author[2]{\fnm{Jonathan S.} \sur{White}}\email{jonathan.white@psi.ch}

\author[3]{\fnm{Daniel} \sur{Mayoh}}\email{D.Mayoh.1@warwick.ac.uk}

\author[3]{\fnm{Geetha} \sur{Balakrishnan}}\email{g.balakrishnan@warwick.ac.uk}

\author[4]{\fnm{Diego} \sur{Alba Venero}}\email{diego.alba-venero@stfc.ac.uk}

\author[1]{\fnm{Peter D.} \sur{Hatton}}\email{p.d.hatton@durham.ac.uk}

\affil*[1]{\orgdiv{Department of Physics}, \orgname{Durham University}, \orgaddress{\street{South Road}, \city{Durham}, \postcode{DH1 3LE}, \country{United Kingdom}}}

\affil[2]{\orgdiv{Laboratory for Neutron Scattering and Imaging (LNS)}, \orgname{Paul Scherrer Institut (PSI)}, \orgaddress{\postcode{CH-5232}, \city{Villigen}, \country{Switzerland}}}

\affil[3]{\orgdiv{Department of Physics}, \orgname{University of Warwick}, \orgaddress{\street{Street}, \city{Warwick}, \postcode{CV4 7AL}, \country{United Kingdom}}}
\affil[4]{\orgdiv{ISIS Neutron and Muon Source}, \orgname{Rutherford Appleton Laboratory}, \city{Didcot}, \postcode{OX11 0QX}, \country{United Kingdom}}


\abstract{Magnetic skyrmions are vortex-like, swirls of magnetisation whose topological protection and particle-like nature have suggested them to be suitable for a number of novel spintronic devices. One such application is skyrmionic computing, which has the advantage over conventional schemes due to the amalgamation of logic calculations and data storage. Using small-angle neutron scattering from \ce{Cu2OSeO3}, and applying electric and magnetic fields, we find that the direction of the skyrmion-coexisting conical states can be manipulated by varying the electric field, and explain this using a free energy approach. Our findings unlock the prospect of creating a number of skyrmion devices which may constitute part of a skyrmion computer, as the direction of a skyrmion within a nanosized racetrack can be manipulated into different channels by controllably changing the direction of the localised conical state. We provide time-dependant micromagnetic simulations to demonstrate such a device: a skyrmion double transistor.}




\maketitle

The current data revolution is being driven by rapid developments within the field of electronics, which has been accelerating since the Nobel-prize-winning discovery of the transistor \cite{Bardeen}. However, the ever-growing demand for higher processing speeds together with more compact data storage has lead to significant environmental concerns due to inevitable increases in energy consumption. One solution is the field of spintronics, which utilises unique magnetic and spin phenomena at the nanoscale to reduce both the input power and device footprint as well as increase data speeds compared with conventional electronic devices. The development of hard drives after the discovery of giant magnetoresistance epitomises the impact of the field and also culminated in a Nobel prize \cite{Baibich}. Since then, spintronics has exploded with the investigation of novel magnetic textures that host exotic physical properties being a promising avenue and is receiving intense research efforts. \\

Magnetic skyrmions are one such avenue of spintronics research. They are vortex-like whirls of magnetisation, with a swirling structure defined by:

\begin{equation}
\label{winding}
    N = \frac{1}{4\pi}\int \mathbf{\hat{n}}\cdot(\frac{\partial\mathbf{\hat{n}}}{\partial x}\times\frac{\partial\mathbf{\hat{n}}}{\partial y}) \; \textnormal{d}^2r,
\end{equation} \\

where the integrand measures the solid angle between neighbouring spins, $\mathbf{\hat{n}}$. The skyrmion number, $N$, takes on integer values if the total solid angle spanned by the spin-texture wraps completely around a unit sphere. This wrapping grants skyrmions a so-called topological stability, granting them protection against continuous deformations that may induce transitions into magnetic states of different winding number. Magnetic skyrmions were first discovered in the chiral helimagnet \ce{MnSi} \cite{Muhlbauer2009}, many years after the observation of one-dimensional helical and conical states within the same material \cite{ishikawa1977}. They have since been found in other bulk chiral magnets, including \ce{FeGe} \cite{Yu2011}, \ce{Fe_{0.5}Co_{0.5}Si} \cite{Yu2010}, Co-Zn-Mn alloys \cite{Tokunaga2015, Ukleev2021} and \ce{Cu2OSeO3} \cite{Seki_CSO}. The hunt for skyrmions and the subsequent investigation of their physical properties has been driven by the prospect of using them within a vast array of possible spintronic devices, with proposals ranging from energy-efficient non-volatile racetrack memory \cite{Fert_2013, Je2021, Zhang2020, Tomasello2014}, neuromorphic, stochastic and reservoir computing schemes \cite{Song2020, Zazvorka2019, ZhangH2020, Pinna_2020}, as well as microwave resonance detectors and logic-gates for direct-skyrmion computing \cite{Okamura2013, Luo2018,  Luo2021}. For these spintronic devices to be realised, discovering and quantifying mechanisms for controlling the direction of motion for individual skyrmions are essential. \\

In this article, we show that the application of an electric field on the Mott-insulator, \ce{Cu2OSeO3}, can cause macroscopic reorientations in the propagation direction of the skyrmion-coexisting conical states. The application of the $E$-field leaves the orientation and the stability of the skyrmions unchanged. We find that the conical deflection angle is linear with both sign and magnitude of the applied field, and explain this observation using a continuum-level theory. This direct manipulation opens up the possibility of using conical states to control the motion of skyrmions within skyrmionic devices, due to the inherent repulsion between two localised, well-defined magnetic textures \cite{Brearton_2020}. 
Such a mechanism for skyrmion motion control would be highly suitable for nanostructured, single-crystalline skyrmion hosts which provide an attractive alternative to the well-studied skyrmion-hosting synthetic magnetic multilayers due to an inherent decrease in the number of dissipative skyrmion pinning sites \cite{Zeissler2017}. Furthermore, this mechanism opens up the possibility of a lossless $E$-field control of magnetism within insulators. To highlight the potential of our findings, we perform time dependant micromagnetic simulations as a proof of concept for a 2D skyrmion double transistor, which is pertinent for the development of skyrmion computing.\

\section*{Spin Textures in \ce{Cu2OSeO3}}

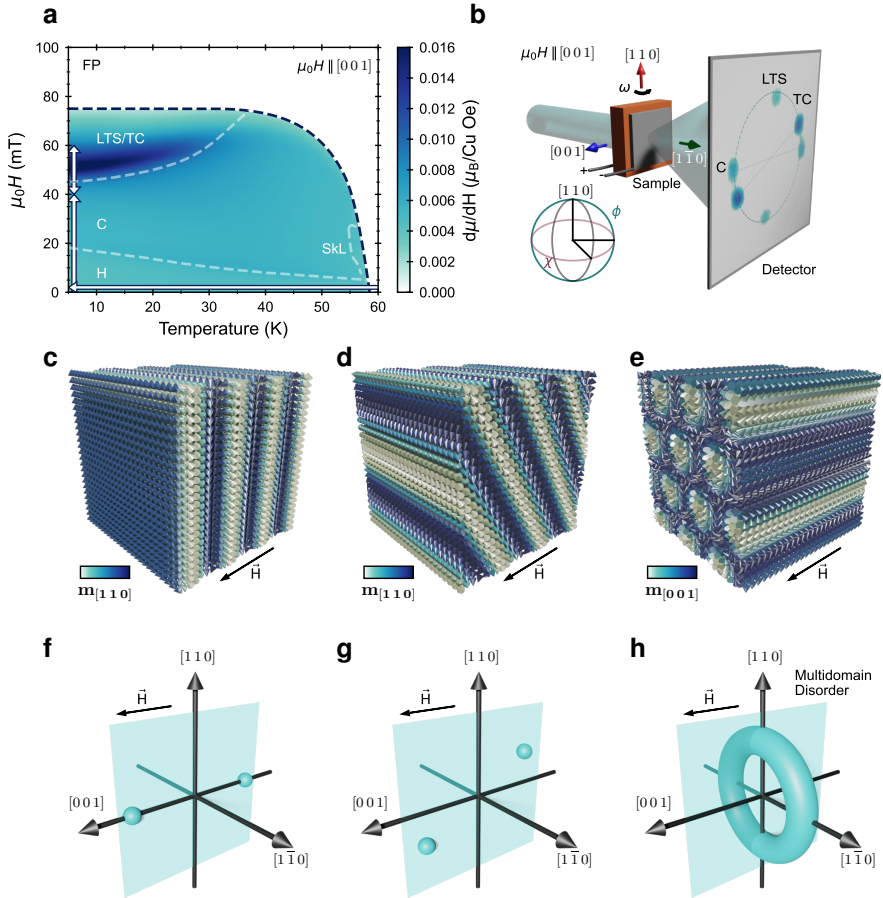
\begin{figure}
\begin{center}
\centering
\includegraphics[width=\linewidth]{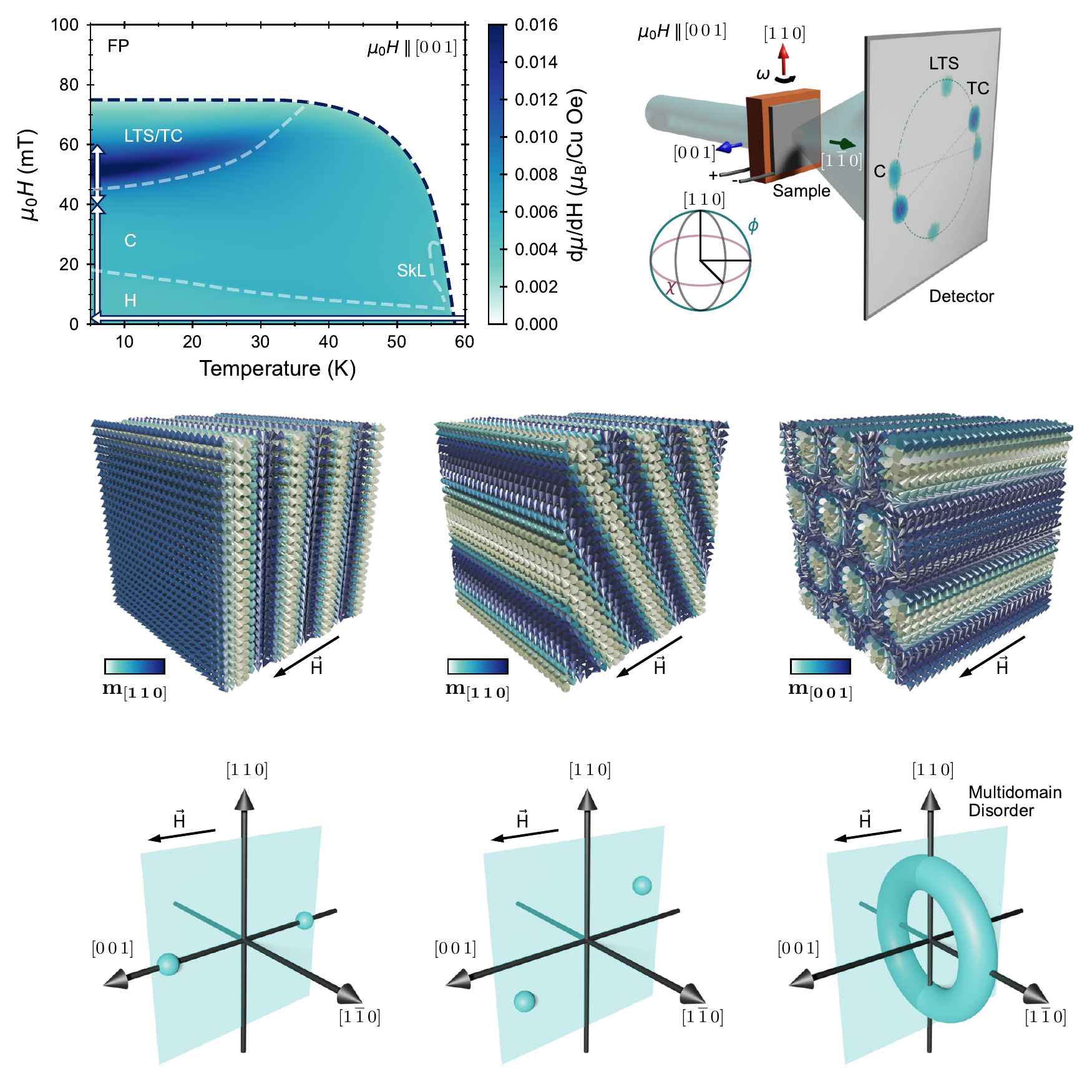}
\put(-320,340){\textbf{a}}
\put(-160,340){\textbf{b}}
\put(-320,210){\textbf{c}}
\put(-210,210){\textbf{d}}
\put(-100,210){\textbf{e}}
\put(-320,100){\textbf{f}}
\put(-210,100){\textbf{g}}
\put(-100,100){\textbf{h}}
\caption{ \centering \textbf{a}, $H-T$ magnetic phase diagram of \ce{Cu2OSeO3} for H $\parallel$ \hkl[001] which hosts 6 magnetic phases: helical (H), conical (C), A-phase skyrmions (SkL), field-polarised (FP), tilted conical (TC) and low-temperature skyrmions (LTS). \textbf{b}, SANS scattering geometry: incoming neutrons parallel to the \hkl[1-10] direction with the applied magnetic field perpendicular to the neutron beam, parallel to \hkl[001]. Silver paste contacts on the surfaces apply a uniform electric field along the \hkl[1-10] direction. \textbf{c-e}, Real space magnetic textures of the conical phase (\textbf{c}) tilted conical phase (\textbf{d}) and ordered skyrmions (\textbf{e}, LTS are disordered due to multiple domains) with the direction of the magnetic field ($\mathbf{{H}}$) shown. \textbf{f-h}, The corresponding reciprocal space patterns of the phases in \textbf{c-e} observed on the detector with the plane indicating the SANS diffraction condition. A number of disordered skyrmionic domains form within the LTS state, such that the diffraction pattern azimuthally broadens to form a ring within reciprocal space.}
\label{f1}
\end{center}
\end{figure}

The presence of spin-orbit coupling and the lack of centrosymmetry in \ce{Cu2OSeO3} allows a non-vanishing Dzyaloshinskii-Moriya interaction (DMI, magnitude $D$), which is energy-minimised by a perpendicular arrangement of neighboring magnetic moments. The DMI competes with the exchange interaction (magnitude $A$), to stabilise long wavelength helical states ($\lambda = \frac{A}{D} \approx 62$ nm) below the Curie temperature of 58 K. This competition also allows a number of other incommensurate spin textures which are closely spaced in free energy. As shown in the magnetic phase diagram in Fig.~\ref{f1}a, transitions between textures can be induced via a magnetic field, which typically causes a transition from a multi-domain helical (H) into a single-domain conical state (C). However, at temperatures towards the ordering temperature ($T \approx$ 56 K) and far below ($T < $ 25 K), the magnetic field drives the formation of magnetic skyrmions that are stabilised by thermal fluctuations and magnetic anisotropies respectively \cite{Muhlbauer2009, Chacon_2018}. At low temperatures, weaker magnetic interactions that are higher-order in the spin-orbit coupling, such as magnetocrystalline anisotropies (MCA) and anisotropic exchange interactions (AEI), become more significant and allow transformations into novel magnetic states \cite{Moody2021}. These states can be investigated using small-angle neutron scattering (SANS), which gives a quantitative measurement on the populations, ordering and periodicity of each incommensurate magnetic state. In our experimental geometry (see Methods and Fig.~\ref{f1}b), we are able to separate the diffracted intensity from the conical (C), tilted conical (TC) and low temperature skyrmions (LTS) due to the spatial separation of their respective diffraction peaks on the detector. The real-space configuration of these spin-textures is shown in Fig.~\ref{f1}c-e. In (c), the conical state is characterised by both the conical propagation direction (wavevector) and the canting of spins along the field direction by an angle $\theta$ to minimise the Zeeman energy. The TC state in (d) is formed by the interplay between the AEI and MCA, which causes the wavevector to deviate from the direction of applied magnetic field along a \hkl[100] towards the \hkl<111> directions by an angle $\phi$, splitting into four degenerate domains \cite{Qian2018}. The swirling nature of the LTS is shown in (e), and are stabilised within a plane perpendicular to a field along the \hkl[001] due to cubic anisotropy, with no other orientational preference \cite{Chacon_2018, Halder2018}. \\ 

Our SANS experiment can detect the periodic components of the magnetic textures that are perpendicular to the incoming neutron beam (within the blue plane), schematics of the spatially dependant Fourier transforms of the C, TC and LTS states are shown below their real-space counterparts in Fig.~\ref{f1}f-h. As shown, the single-frequency sinusoidal magnetisation profile of the C and TC states allows only a single pair of spots, whose locations give information about the direction of propagation and wavelength of the their respective spin-textures. On the other hand, the LTS exhibit magnetisation modulations perpendicular to the magnetic field direction and form within spatially separated domains at random orientations,  which gives a ring-like shape within frequency space. The LTS ring meets the diffraction condition only when it intersects the diffraction plane, which we detect as two spots along the vertical direction. In this geometry, the coexistence between various states is easily observed, since the states are remarkably similar in free energy \cite{Chacon_2018}. This coexistence is easily achieved in real samples and has the potential to be utilised within skyrmionic devices, as these states occupy local minima within the magnetic Hamiltonian, and as such have repulsive interactions \cite{Brearton2021}. Furthermore, the application of an electric field introduces an additional term within the free energy which has the potential to break the near-degeneracy of the magnetic states. Here we perform the first investigation of the electric field effects on these recently discovered low-temperature magnetic states and reveal their potential implications for use within skyrmionic devices. 

\section*{$E$-Field SANS}

\begin{figure}
\begin{center}
\centering
\includegraphics[width=1\textwidth]{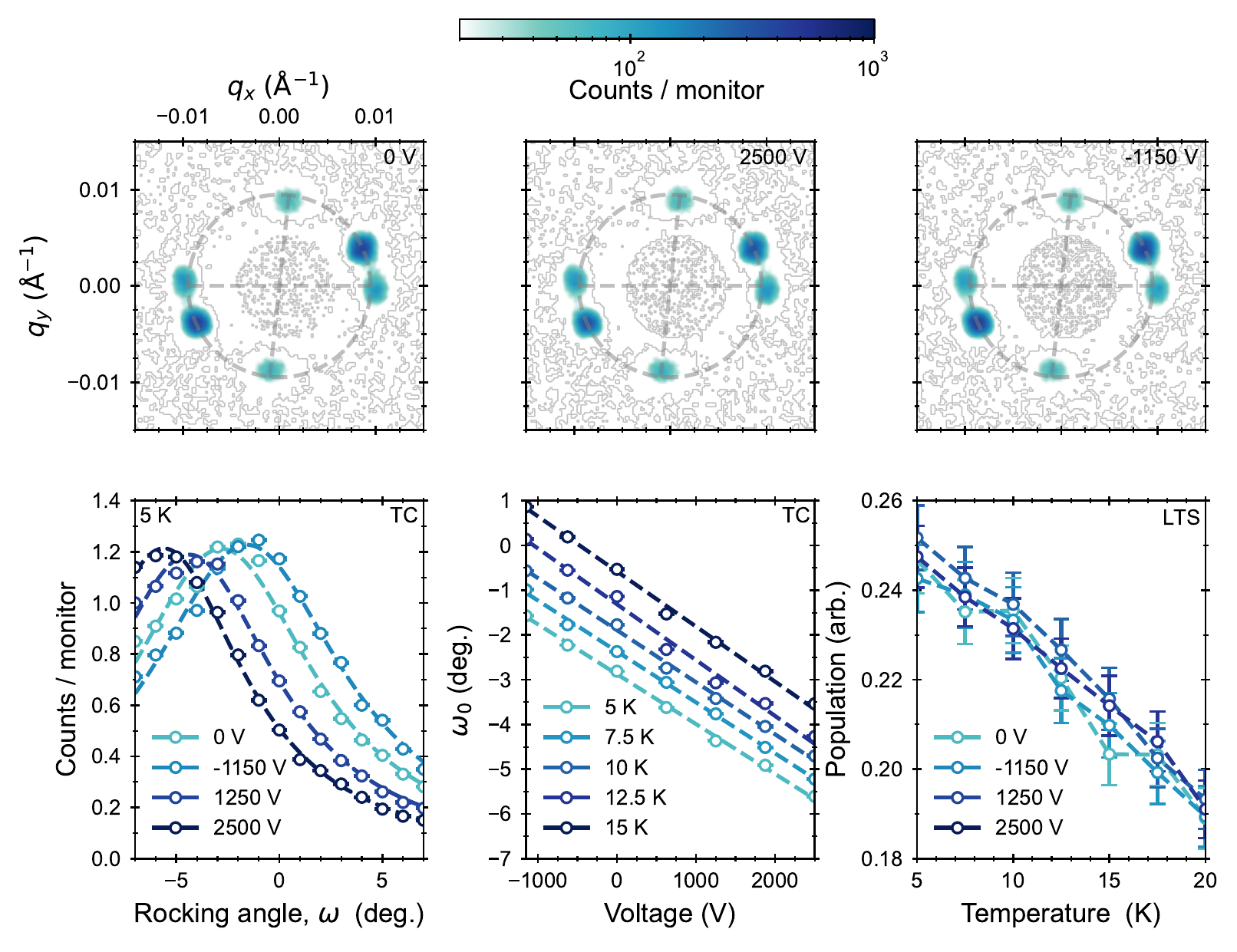}
\put(-260,210){\footnotesize LTS}
\put(-237,195){\footnotesize TC}
\put(-232,180){\footnotesize C}
\put(-310,220){\textbf{a}}
\put(-205,220){\textbf{b}}
\put(-98,220){\textbf{c}}
\put(-310,130){\textbf{d}}
\put(-205,130){\textbf{e}}
\put(-98,130){\textbf{f}}
\caption{\textbf{a-c}, Rocked SANS patterns in a 40 mT magnetic field at 5 K, after 40 field oscillations within an electric field of 0 V (\textbf{a}), 2500 V (\textbf{b}) and -1150 V (\textbf{c}) respectively. \textbf{d}, Rocking scans of the tilted conical intensity taken at a series of electric fields, showing a shift in the peak proportional to the electric field. \textbf{e}, The linear dependence of the peak center ($\omega_0$) for applied voltage for a selection of temperatures. \textbf{f}, Temperature dependence of the LTS state for different electric fields.}
\label{f2}
\end{center}
\end{figure}
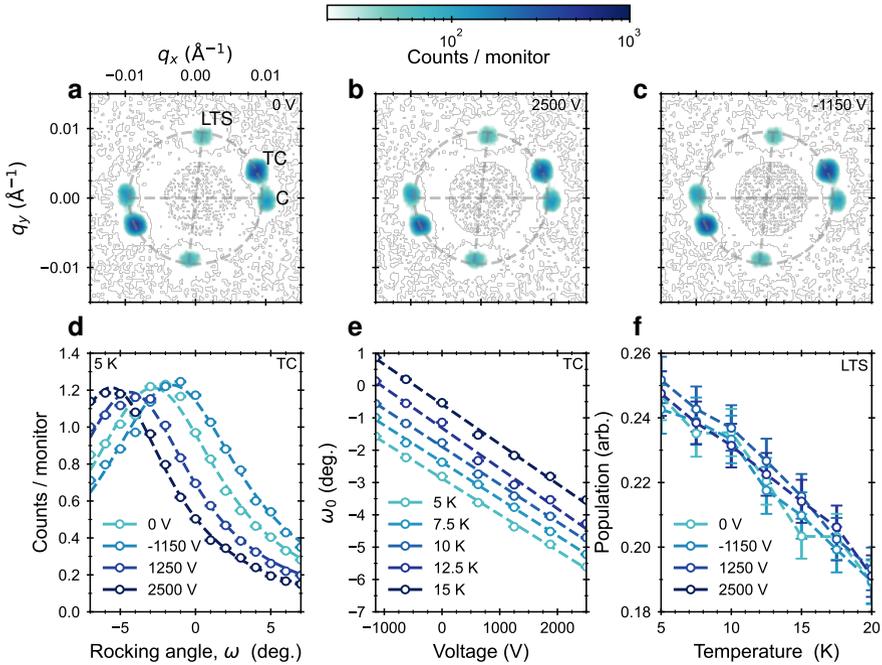

To quantify the effect of an applied electric field on the LTS and TC states in \ce{Cu2OSeO3}, we employed a zero-field-cooled procedure (See Fig.~\ref{f1}a and Methods) to obtain consistent states at low temperatures for which any differences depend only on the magnitude and sign of the applied electric field. As shown in the summed rocking scans in Fig.~\ref{f2}a-c, we observe no differences in the magnitude of the wavevectors or the diffracted intensities for any of the three coexisting magnetic phases, despite being at the limits of the experimentally accessible $E$-fields with $E=$ 0, 2.5 and -1.150 \SI{}{\kilo\volt\per\milli\meter} respectively. \\

However, by investigating the $E$-field dependent behaviour of the rocking scans for the TC peaks shown in Fig.~\ref{f2}d, we find that the peak of the rocking curve ($\omega_0$) varies as a function of the applied field, shifting towards larger (smaller) angles for negative (positive) electric fields respectively. The shifting of $\omega_0$ reveals that the propagation direction of the TC spin texture deviates within a finite $E$-field away from the $E = 0$ alignment, in a direction along the equatorial polar angle, $\chi$ (parallel with neutron beam, see Supplementary Information Fig.~1 for angular definitions). This behaviour is found at a variety of applied $E$-fields and temperatures, as shown in Fig.~\ref{f2}e, with different temperatures offset in $\omega$ for clarity. Here, $\omega_0$ varies linearly for both positive and negative applied $E$-fields, at all temperatures investigated.  \

In contrast, no magnetoelectric effects on the LTS are observed. Unlike previous studies of $E$-field effects on skyrmions in the A-phase, which demonstrated the $E$-field control of the SkL orientation and phase stability \cite{White_2012,white_2014, Kruchkov_2018, White2018, Wilson_2019}, the direction of magnetic modulations of the LTS remain firmly fixed in the plane spanned by the \hkl[100] and \hkl[010] directions, resulting in a zero net electric polarisation due to each skyrmion carrying a cancelling electric quadrupole \cite{Seki2012}. This lack of coupling can be seen in the temperature dependence of the LTS intensity, Fig.~\ref{f2}f, which shows identical LTS populations and ordering within statistical limits (see Supplementary Information Figs~2,3.). Similar population behaviour is found for the TC, which displays a similar peak height in Fig.~\ref{f2}d but the limited angular range hinders a thorough characterisation. \

The ability to reliably control the orientation of the TC state using an $E$-field whilst not altering the stability of the co-existing skyrmions provides new functionality paradigms that may find use in skyrmionic devices. In what follows, we will show that this mechanism could be used to control skyrmion motion by dynamically altering the propagation direction of the TC state, such that it acts as a switchable barrier that is able to deflect a skyrmion under motion along different trajectories. In order to develop such a device, it is essential to engineer materials which host the specific anisotropic and magnetoelectric properties. Here we develop the theoretical requirements for such a device below.\

\section*{Microscopic Origin of the $E$-field Induced Tilted Conical Reorientation}

The $E$-field-induced reorientations of the tilted conical wavevector within \ce{Cu2OSeO3} highlights the remarkable anisotropic and magnetoelectric properties of the material.  These properties cause particular magnetic textures within \ce{Cu2OSeO3} to generate electric polarisation in finite magnetic field via the metal-ligand hybridisation mechanism, shown to be induced by a relativistic spin-orbit interaction \cite{Arima2007, Yang2012}. Since the tilted conical state is incommensurate with a periodicity much greater than the chemical unit cell, we explain our observations using a mean-field theory. According to the cubic crystal symmetry, the continuum form of the electric polarisation, $\mathbf{P}$, is given by \cite{Mochizuki2015, Mochizuki2013, Liu2013}:

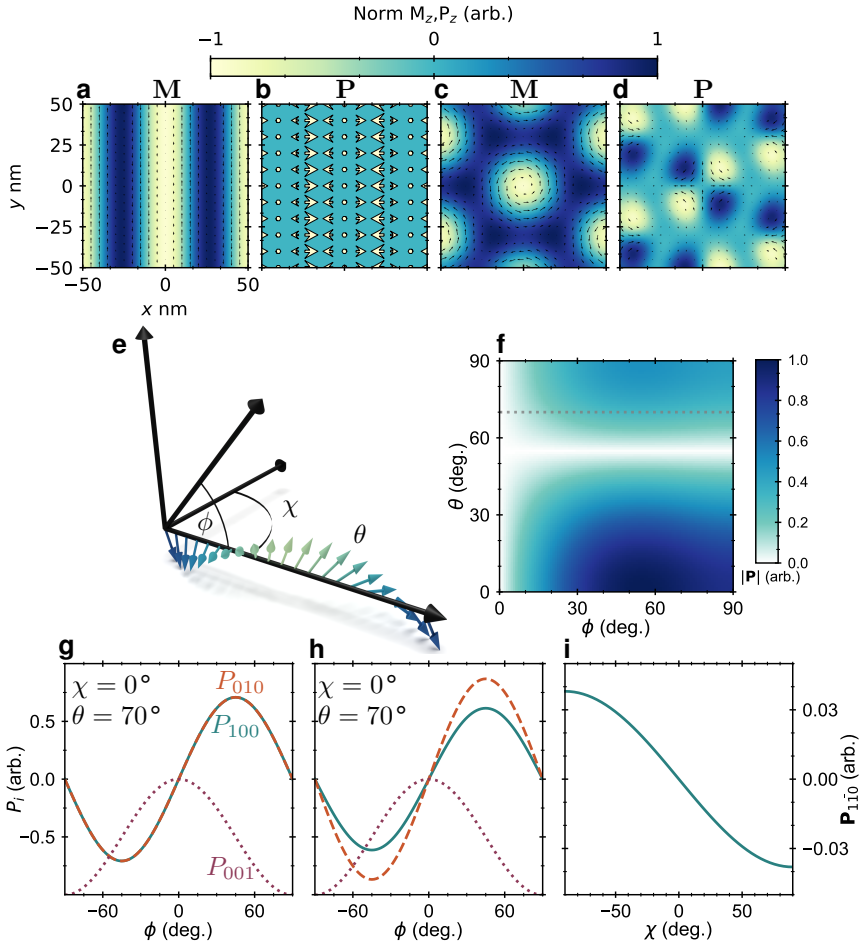
\begin{figure}
\begin{center}
\begin{overpic}[width=0.99\linewidth]{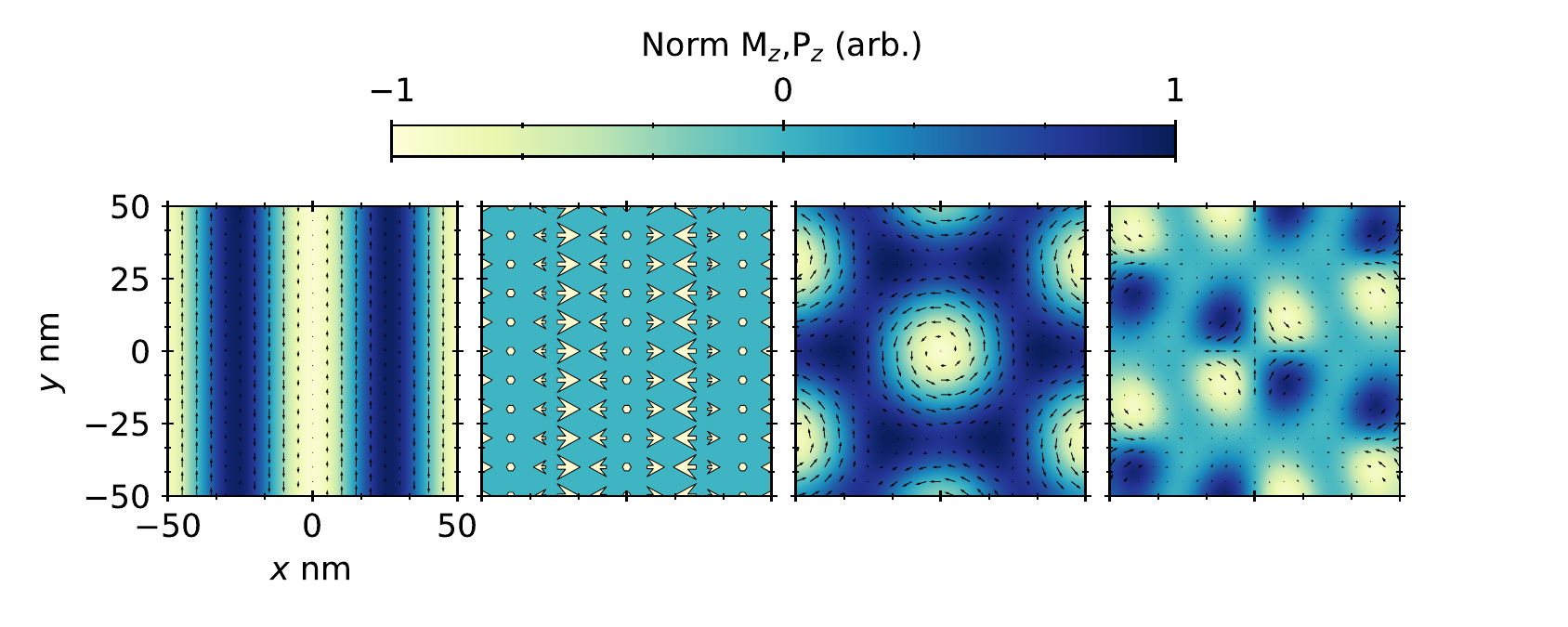}
\put(10, 27.5){\textbf{a}}
\put(30, 27.5){\textbf{b}}
\put(50, 27.5){\textbf{c}}
\put(70, 27.5){\textbf{d}}
\put(18.5, 27.5){$\mathbf{{M}}$}
\put(39, 27.5){$\mathbf{{P}}$}
\put(58.5, 27.5){$\mathbf{{M}}$}
\put(79, 27.5){$\mathbf{{P}}$}
\end{overpic}
\begin{overpic}[width=0.99\linewidth]{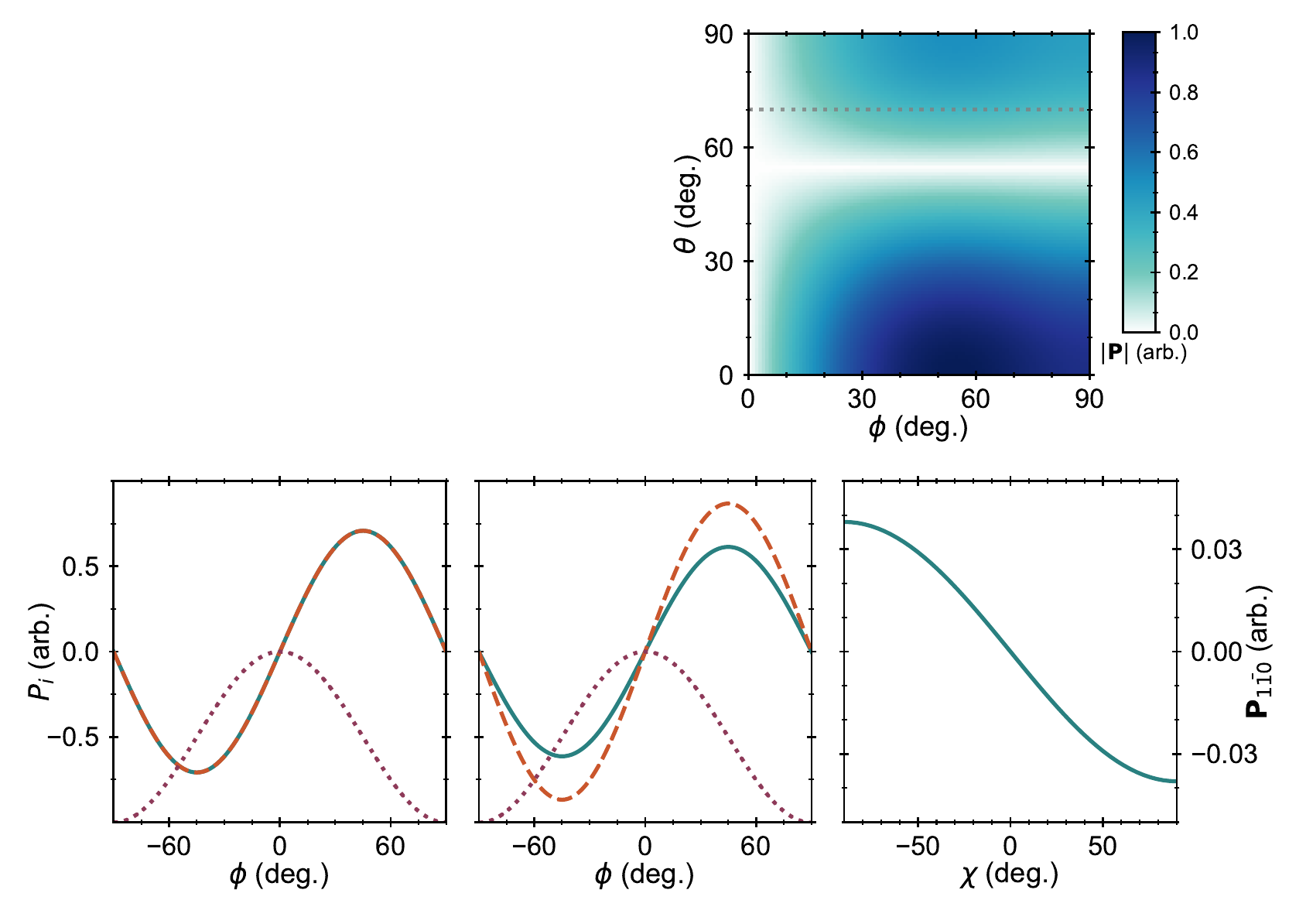}
\put(16, 35){\includegraphics[trim={5cm 5cm 5.5cm 2.7cm}, clip, width=0.35\linewidth, grid]{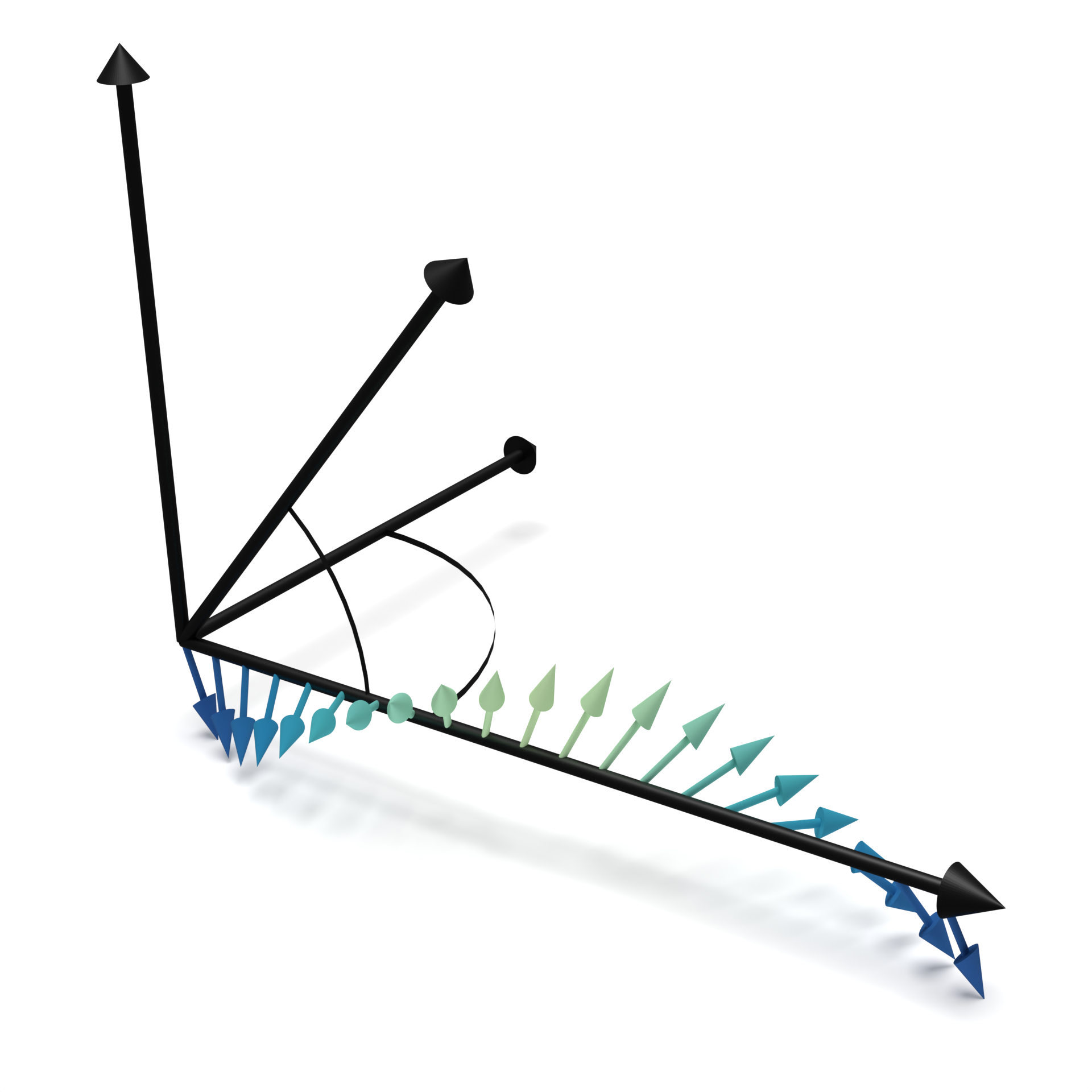}}
\put(14,69){\textbf{e}}
\put(33,51){$\chi$}
\put(23.5, 49){$\phi$}
\put(41, 47.5){$\theta$}
\put(57,69){\textbf{f}}
\put(9.5,30.5){$\chi = 0$\degree}
\put(9.5,27){$\theta = 70$\degree}
\put(8,35){\textbf{g}}
\put(36,34.5){\textbf{h}}
\put(37,30.5){$\chi = 0$\degree}
\put(37,27){$\theta = 70$\degree}
\put(24.5,10){\textcolor{palat}{$P_{001}$}}
\put(25.5,31){\textcolor{myorag}{$P_{010}$}}
\put(25,26){\textcolor{myblue}{$P_{100}$}}
\put(64.5,34.5){\textbf{i}}
\end{overpic}
\caption{\textbf{a}, Magnetisation of a helical spin-texture with wavevector along the \hkl[100] ($x$) direction. \textbf{b}, The electric polarisation induced by the helical state in (a). \textbf{c}, Magnetisation of a skyrmion lattice with periodic components confined within the \hkl[100]-\hkl[010] ($x$-$y$) plane and concomitant polarisation in (\textbf{d}) which shows a cancelling electric quadrupole. \textbf{e}, Schematic of a titled conical state with conical angle $\theta$ ($\theta = 90, 0\degree$ for helical/field-aligned state respectively). Polar-angles of $\phi, \chi$ are used to describe the orientation of the TC state. \textbf{f}, Map of total polarisation, $\mid\textbf{P}\mid$, as a function of $\theta$ and $\phi$ for a particular equatorial angle $\chi=0$. \textbf{g}-\textbf{h}, The individual components of the $\textbf{P}$-vector as a function of equatorial angle $\phi$ with $P_{100}, P_{010}, P_{001}$ being blue solid, brown dashed and purple dotted respectively and with $\chi = $ 0, 5 degrees for (\textbf{g}) and (\textbf{h}) respectively in agreement with experimental observations. The $\phi$ dependence follows the path given by the grey dashed line in (\textbf{a}). \textbf{i}, Induced polarisation along the \hkl[1 -1 0], $P_{100}-P_{010}$, as a function of equatorial angle $\chi$ for a fixed $\phi = \ang{30}$, taken from experimental observations.}
\label{f3}
\end{center}
\end{figure}

\begin{equation}
    \mathbf{P} = (P_{x}, P_{y}, P_{z}) = \lambda^c_{me}(m_ym_z, m_xm_z, m_xm_y)
\label{pol}
\end{equation}

where $\{x, y, z\}$ define a basis following the \hkl<100> crystal directions. The material parameter, $\lambda^c_{me}$, is the magnetoelectric coupling constant and is related to the strength of the spin-orbit coupling interaction. \\

As seen in Eq.~\ref{pol}, the electric polarisation is determined by the moment directions that constitute a particular spin texture, and we further show this dependence in Fig.~\ref{f3}a-d. For a magnetic helix directed along the \hkl[100] direction (a), the resultant polarisation (b) appears with twice the helical frequency and aligns along the propagation vector of the helix. In (c), we show the magnetisation of a skyrmion lattice with periodic components confined within the \hkl[100], \hkl[010] plane (in agreement with LTS) we find that each skyrmion possess an electric quadrupole moment in (d). For both of these spin-textures, integrating across the magnetic unit cell results in net-zero polarisation, and as such little electric field dependence is measured. However, magnetic structures which deviate away from high-symmetry crystal directions may induce a finite polarisation which doesn't vanish across a magnetic unit cell. \\

One example of such a magnetic state with a non-vanishing polarisation is the tilted conical state. The orientational dependence ($\phi$) of the electric-polarisation generated by a tilted conical state for any conical angle ($\theta$, see Fig.~\ref{f3}e and Supplementary Fig.~S1 for angular definitions, and Methods for derivation) is shown in Fig.~\ref{f3}f. For $\phi=0$, the wavevector, $\mathbf{q}$, lies along the \hkl[001] direction, and we find there is no polarisation induced for any $\theta$. However, even at minor tilt angles where $\phi>\ang{10}$, the tilted conical states induce a significant electric polarisation for a range of conical angles. The vectorial nature of the polarisation as a function of $\phi$, is shown in Fig.~\ref{f3}g and h. This dependence follows the grey-dashed line in Fig.~\ref{f3}a, with $\theta = 70\degree$ in agreement with the derivation in Methods. For the case where the equatorial ($\chi$) angle is zero (Experimentally realised for $E = 0$ V), the polarisation along the \hkl[100] and \hkl[010] crystal directions show an equal, antisymmetric behaviour about $\phi=0$, in contrast to $P_{001}$. In this configuration, the polarisation components would then cancel or lie orthogonal to the applied $E$-field along the \hkl[1 -1 0], and the magnetoelectric coupling term would not contribute to the overall magnetic free energy.

However, when an equatorial rotation, $\Delta\chi$ is introduced, the degeneracy between $P_{100}$ and $P_{010}$ is broken and a non-vanishing polarisation ($\mathbf{P_{1\bar{1}0}}$) is induced, see Fig.~\ref{f3}h. Here we choose $\chi=\ang{5}$ in agreement with the maximum deflection observed in Fig.~\ref{f2}. Due to its antisymmetric properties, this polarisation direction changes sign depending on the value of $\phi$, akin to reversing the $E$-field polarity. This change of sign can further be seen in Fig.~\ref{f3}i, which shows the value of $\mathbf{P_{1\bar{1}0}}$ as a function of $\chi$ for the experimentally determined value of $\phi=\ang{30}$. This deflection of the TC wavevector aloing the $\chi$ direction, when in the presence of the electric field, lowers the free energy of the system via a finite magnetoelectric coupling. However, this deflection causes an increase in anisotropy energy. By expanding the free energy density \cite{Qian2018, Moody2021, Bak1980}, integrating over a conical period and differentiating with respect to $\chi$ (see Methods and Supplementary Information [URL to be inserted by the editor] for full derivation), we find the tilted conical deflection along $\chi$ direction for small $E$ to be: 

\begin{equation} \label{e3}
    \frac{\partial \chi}{\partial E} \approx \frac{\lambda^c_{me}\cos\phi f(\theta)}{\gamma q^2 \sin^2\theta\sin^3\phi+K\sin^4\phi g(\theta)},
\end{equation} \\

 where $\gamma$, $K$ are the AEI and MCA constants respectively. $g(\theta), f(\theta)$ are trigonometric functions of conical angle only. The linear dependence of the $\chi$-deflection with respect to the applied $E$-field in Eq.~\ref{e3}, matches the relationship found in experimental data in Fig.~\ref{f2}e, showing that the magnetoelectric coupling is less significant than the anisotropic magnetic interactions. The magnitude of the $E$-field which reorients the TC direction can be reduced by engineering systems with a large coupling constant $\lambda^C_{me}$, or by minimising the anisotropic constants. However, reducing the value of $K$ and $\gamma$ destabilises the TC texture \cite{Qian2018, Moody2021}, which needs to remain a local minima in the free energy landscape to act as a barrier state within a device setting.

\section*{Skyrmionic Transistor}

\begin{figure}
\begin{center}
\centering
\includegraphics[width=1\linewidth]{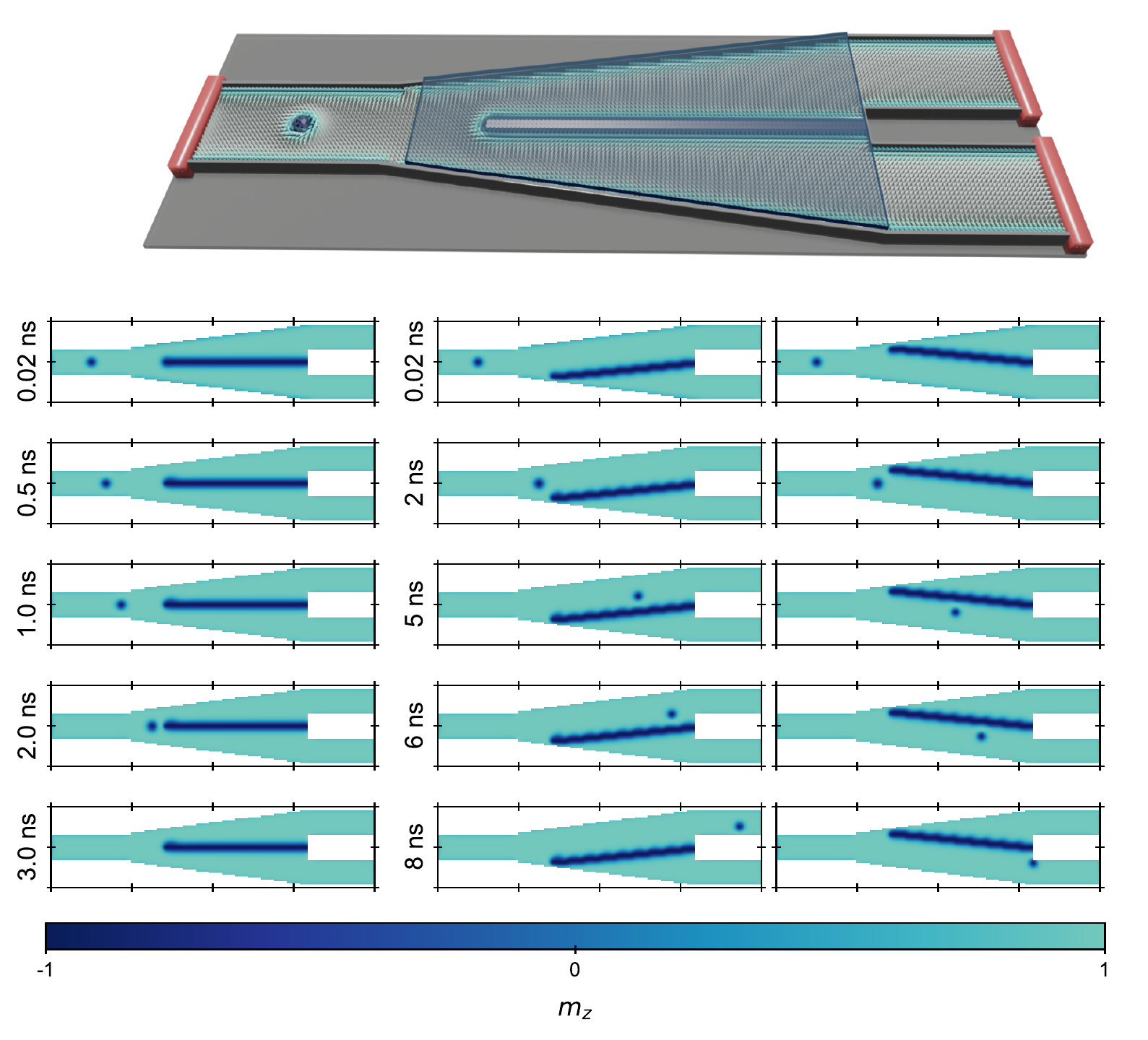}
\put(-330, 310){\textbf{a}}
\put(-330, 220){\textbf{b}}
\put(-215, 220){\textbf{c}}
\put(-115, 220){\textbf{d}}
\caption{\textbf{a}, Example schematic of a skyrmion double transistor with current contacts on either terminal (orange) and the $E$-field contacts illustrated as the blue plate above the division. \textbf{b}, Skyrmion double transistor with the barriar state aligned in the neutral state illustrated as a function of time. The skyrmion propagates along the track until it is annihilated by the localised barrier texture. \textbf{c} and \textbf{d}, Skyrmion double transistor aligned as one of the two active states such that the localised barrier texture is oriented to mimic the effect of an applied $E$-fieldon the TC state. In each case the skyrmion is repelled by the barrier forcing its propagation along one of the two possible paths. Movies of the skyrmion motion can be found here: [URL to be inserted by the editor]}
\end{center}
\label{f4}
\end{figure}
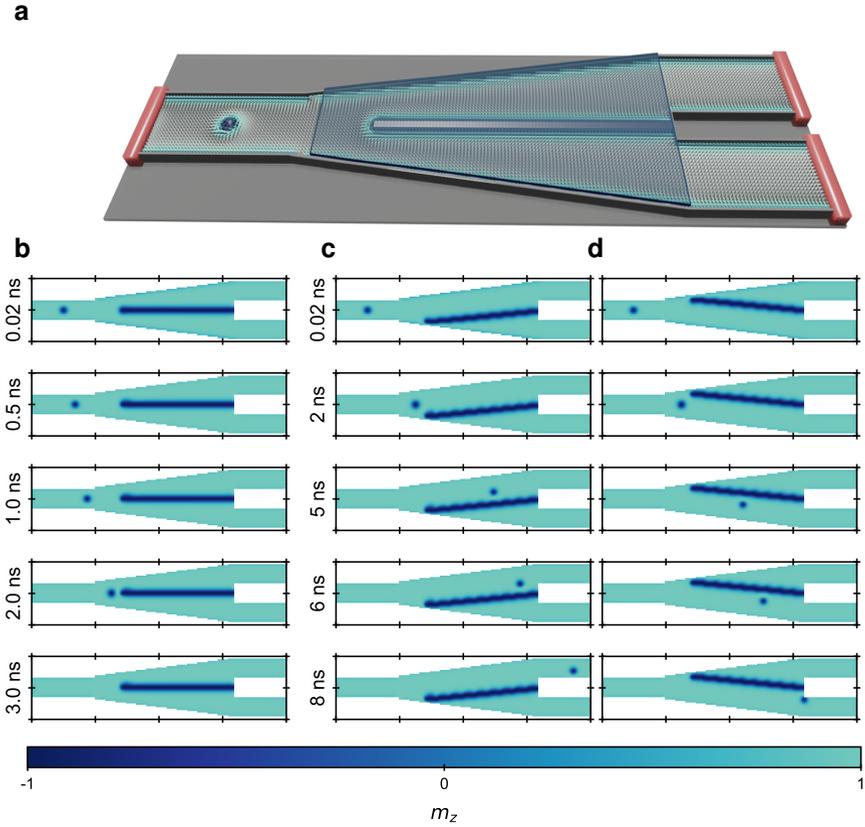

We have demonstrated that by using an $E$-field, we can deterministically control the propagation direction of the TC texture. This presents an interesting opportunity for skyrmionic devices by utilising the fact that well-behaved magnetic states, that are each local minima of the magnetic Hamiltonian (e.g. isolated LTS and localised TC states, shown theoretically to exist \cite{Leonov2019}), by definition have repulsive interactions \cite{Brearton_2020}. This inherent repulsion between coexisting states allows a novel mechanism to directly manipulate the direction of moving skyrmions within devices settings.  For example, individual skyrmions within a racetrack can have their trajectories altered through the interaction with a localised conical state. The skyrmion motion alters according to the orientation of a localised conical state, which is dependent on the magnitude and direction of an applied $E$-field. To prove the feasibility of the physics behind such a mechanism, we used time-dependant micromagnetics (see Methods) to simulate a skyrmionic double transistor and give an overview of the simulation in Fig.~\ref{f4}a, together with movies [URL will be inserted by the Editor]. \\

In our simulations, an isolated skyrmion is driven downstream towards the two possible exits, guided by edge states \cite{Leonov2017}. A number of mechanisms have been proposed to induce isolated skyrmion motion, including thermal and magnetic field gradients \cite{Lin2013, Zhang2018, Brearton2021}. However, despite \ce{Cu2OSeO3} being an insulator, we decided to use the only available micromagnetically-implemented method of using a spin-polarised current ($j_x =$ 100 \SI{}{\ampere\per\meter\squared}), which drives the skyrmions via spin transfer torque \cite{Iwasaki2013}. In reality, one could use an alternative driving method, engineer a two-layer device consisting of conducting multilayer that is strongly coupled to an insulating region, or add insulating contacts across the device that would allow the creation of a temporary electric field to set the direction of a pinned conical state within a conductive sample via an instantaneous magnetoelectric coupling. \\

Initially, we relax the system to form an isolated skyrmion, together with a strip of negative magnetisation to act as the barrier state in place for the tilted cone, since the 2D projection of a tilted cone with a wavevector component along $z$ is non-physical. We give the barrier state three different starting angles to mimic the effect of using an electric field with $E<0, E=0$ and $E>0$, analogous with our experimental results. For $E=0$ (Fig.~\ref{f4}b) the skyrmion is forcefully driven towards the barrier state, before eventually colliding and annihilating with the barrier state, comparable to the edge-state destruction of skyrmions \cite{david2017}. However, in Figs.~\ref{f4}b and c, the inherent repulsive interaction between the barrier state and the skyrmion causes an alteration in the skyrmion's trajectory. This alteration directs the skyrmion to move towards one of two possible exits. Depending on the angular direction of the barrier state, the skyrmion arrives at the relevant exit gate at different times due to the the skyrmion Hall-angle \cite{Zeissler2020}, which arises as a consequence of the skyrmion winding number in Eq.~\ref{winding}, by either aiding/hindering the deflection away from the barrier state. \\

This gives the device three possible output states:  skyrmion deflected down, skyrmion deflected up, and skyrmion annihilated. This output is analogous to an electronic double transistor, which are frequently used in series for the creation of the NOR logic-gate; a universal logic gate that can perform the AND, OR and NOT logic-functions. The practical realisation of such a device would allow for all of the operations required by skyrmion computing, as well as providing a useful method to control the number of skyrmions within other proposed skyrmion computing schemes \cite{Song2020, Zazvorka2019, ZhangH2020, Pinna_2020}. \

\section*{Conclusion}

In conclusion, we have shown that the direction of the skyrmion-coexisting conical states can be controlled by using an applied electric field. We explain the microscopic origin for the $E$-field-induced conical state deflection by using mean-field theory and show that this deflection is linear with the magnitude of $E$. Our findings open up the possibilities of using localised conical states within nanosized skyrmion racetracks as bumper states, which guide the skyrmions towards particular outputs whose outcome can be controlled using an electric field. In particular, we have performed time-dependant micromagnetic simulations of a skyrmion double-transistor device, which is a  key element in realising skyrmion computing and other advanced spintronic devices.

\bibliography{newbib}

\section*{Online Content}

\section*{Methods}
\subsection*{Sample Preparation}
Single crystals of \ce{Cu2OSeO3} were grown from 5 g of thoroughly mixed stochiometric amounts of CuO (99.99\%, Alfa Aesar) and \ce{SeO2} (99.999\%, Arcos Organics) powders using the chemical vapour transport method. The powder along with 2 mg/cc of \ce{TeCl4} transport agent was sealed in an evacuated silica tube with the heating source maintained at 640 \degree C with the sink at 550 \degree C for four weeks. Several single crystals with dimensions of approximately $4 \times 4 \times 4$ mm$^3$ were obtained. For the SANS experiment, one single crystal was shaped using a polishing wheel into a plate with dimensions $3.5\times 4\times 1$ mm$^3$ and a mass of 51.7 mg. The sample was polished such that there two large, parallel ($1\bar{1}0$) faces

\subsection*{Magnetic Property Measurements}
A Quantum Design Magnetic Property Measurement System, MPMS3, superconducting quantum interference device (SQUID) magnetometer was used to investigate the bulk magnetic properties of the sample as a function of both temperature and magnetic-field. The single-crystal used for the SANS measurements was aligned such that $mu_0H\parallel$\hkl[100] and magnetic field-dependant DC magnetic-susceptibility measurements were carried out between 5 K and 60 K in 2 K intervals by zero-field cooling to the temperature set-point and increasing the magnetic field from 0 - 100 mT in 1 mT steps. Numerical derivatives of the resulting magnetisation were calculated to obtain phase boundaries and ordering temperatures with the first derivative plotted in Fig.~\ref{f1}a. See Supplementary Information for additional DC magnetisation measurements.

\subsection*{SANS Measurement}

Small-angle neutron scattering (SANS) measurements were performed at the SANS-I beamline at the Swiss Spallation Neutron Source SINQ, Paul Scherrer Institut, Switzerland, with neutrons of wavelength $\lambda_n$ = 8 \SI{}{\angstrom}, a wavelength spread $\frac{\Delta \lambda_n}{\lambda_n} = $10\%, and a collimator and detector distance of 18 m. The single-crystal was mounted onto a dedicated $E$-field sample holder (see Bartkowiak \textit{et al.})\cite{Bartkowiak2014}, with electrodes attached directly onto the large, flat \hkl(1-10) faces. The sample was placed within a horizontal-field cryomagnet and oriented such that $E\parallel$ $n_0$ $\parallel$ $\hkl[1-10]$, and $H\perp n_0 \parallel$ \hkl[001], see configuration in Fig.~\ref{f1}c. The rocked SANS diffraction patterns were measured by rotating both the sample and cryomagnet together through an angular range such that the diffraction spots completely passed through the diffraction condition. The sum of the resulting rocking scan over an angular range of 14$^\circ$ (7 degrees either side of the origin) are shown in the Fig.~\ref{f2}(a-c). All measurements were taken at temperatures ranging from 5 K to 60 K and with magnetic fields ranging from 0 - 250 mT. Background measurements were carried out in the paramagnetic state at $T=70$ K and in the field-polarised regime at 5 K, 250 mT. The cooling down procedure was from 60 to 5 K, before simultaneously ramping to the desired electric field and applying 40 magnetic field oscillations from 40 mT to 60 mT, see white arrows in Fig.~\ref{f1}a. This procedure allowed a consistent state to be to be obtained at low temperatures.

\subsection*{Mean-Field Theory}

A zero-temperature free energy density expansion for a slowly-varying spin density $\mathbf{m}(\mathbf{r})$ is given by:

\begin{equation} \label{Energy}
\begin{split}
F(\mathbf{r}) &=  D\mathbf{m}(\mathbf{r})\cdot(\nabla\times\mathbf{m}(\mathbf{r})) \\
 &+\frac{1}{2}A[(\nabla m_x)^2+(\nabla m_y)^2+(\nabla m_z)^2] \\
 &+ \frac{1}{2}\gamma[(\frac{\partial m_x}{\partial x})^2+(\frac{\partial m_y}{\partial y})^2+(\frac{\partial m_z}{\partial z})^2]\\
 &+ K(m_x^4 + m_y^4+m_z^4) \\
 &-\mu_0M_s(\mathbf{m}(\mathbf{r})\cdot\mathbf{H}) - \mathbf{P(\mathbf{r})}\cdot\mathbf{E},
 \end{split}
\end{equation}

where $D, A$ is the Dzyaloshinskii-Moriya interaction coefficient and exchange stiffness respectively. The anisotropy constants, $K, \gamma$, are for the 4$^\textnormal{th}$ order magnetocrystalline anisotropy and anisotropic exchange interactions respectively, and are required take on particular values in order to generate a tilted conical state \cite{Qian2018}. $\mathbf{H}$ and $\mathbf{E}$ are the applied magnetic and electric fields respectively, each linearly couple with the reduced magnetisation $\mathbf{m}(\mathbf{r})$ and polarisation $\mathbf{P}(\mathbf{r})$. We use the same polar coordinate basis to define the conical state wavevector in spherical polar coordinates, where $\mathbf{q} = \mathbf{q}(q, \phi, \chi)$, such that:

\begin{equation} \label{e2}
    \mathbf{m}(\mathbf{r})/M_s = \sin\theta(\cos(\mathbf{q}\cdot\mathbf{r})\uvec{e}_1+\sin(\mathbf{q}\cdot\mathbf{r})\uvec{e}_2) + \cos\theta\uvec{e}_3,
\end{equation}

where $\theta$ is the conical angle, and $\{\uvec{e}_n\}$ define three mutually orthogonal basis vectors, such that $\mathbf{q} \parallel \uvec{e}_3$. To match the experimental configuration, we set the equatorial angle ($\chi=\ang{0}$) such that variations in $\phi$ cause $\mathbf{q}$ to rotate about the \hkl[1 -1 0] direction (neutron beam direction), within the plane perpendicular to the neutron beam during the SANS experiment (azimuthal angle on the detector, see Fig. \ref{f1}b). Using $\mathbf{q} = q(\sin\phi\cos\alpha, \sin\phi\sin\alpha, \cos\phi)$, together with an applied electric field along the \hkl[1 -1 0] direction, $E$, inserting this into Eq.~\ref{Energy}, integrating over one conical period to find the average free energy before differentiating with respect to $\alpha$ (See SI for full derivation), we find: 

\begin{equation}\label{dfda}
\begin{split}
    \frac{\partial \bar{F}}{\partial \alpha} &= \gamma \alpha q^2 \sin^2\theta\sin^4\phi \\
                                             &+ E\lambda^c_{me}\sin\phi\cos\phi(1+\alpha) f(\theta)\\
                                             &+K \alpha  \sin^4\phi g(\theta)
\end{split}
\end{equation}

Where the functions $f(\theta) = \frac{\sin^2\theta}{2}-\cos^2\theta$ and $g(\theta) = 24\sin^2\theta\cos^2\theta - 3\sin^4\theta - 4\cos^4\theta$ are dependant on the conical angle only. As shown, the $(1+\alpha)$ term in Eq. (4) causes a non-zero electric field to shift the solutions of the free energy differential to finite values of $\alpha$. These values can be determined by solving Eq.~(4) to obtain $\alpha(E)$, which for small $E$ we find the linear relationship present in the main text: 

\begin{equation}
    \frac{\partial \alpha}{\partial E} = \frac{\lambda^c_{me}\cos\phi f(\theta)}{\gamma q^2 \sin^2\theta\sin^3\phi+K\sin^4\phi g(\theta)}
\end{equation}

Furthermore, the magnitude of the gradient allows us to determine the conical angle $\theta$ providing the material constants are known. At 5 $K$, the material parameters of $\gamma = -6.7\times 10^{-14}$ \SI{}{\joule\per\meter} \cite{Moody2021}, $K = -0.6 \times10^3$ \SI{}{\joule\cubic\meter} \cite{Halder2018} and $\lambda^c_{me} = 5.64\times10^{-27}$ \SI{}{\micro\coulomb\per\meter} \cite{Mochizuki2015} are known, allowing us to use the gradient of the 5 $K$ dataset to determine the conical angle, $\theta = $\SI{70.3(2)}{\degree}. \

\subsection*{Micromagnetic Simulations}
Micromagnetic simulations of the skyrmionic double transistor device were performed with the Ubermag \cite{beg2022} meta-package which uses OOMMF \cite{OOMMF} as the micromagnetic system driver. The simulated system was specified with total dimensions $1000\times 250 \times 5$ nm, using finite difference cells with a volume of 5 nm$^3$, with a number of cell switched off by setting their magnetisation to zero to obtain the desired geometry. We describe the transistor system using the simplest skyrmion-hosting micromagnetic energy functional of a chiral magnet with symmetry class T, which reads:

\begin{equation}
E = \int_V A(\mathbf{\nabla}\mathbf{m}) + D\mathbf{m}\cdot(\mathbf{m}\times\mathbf{m}) - \mu_0M_s \mathbf{m}\cdot \mathbf{H_a} \; \textnormal{d}V
\end{equation}

where $\mathbf{m}$ is the normalised magnetisation, $A=3.5\times10^{-13}$ \SI{}{\joule\per\meter} is the continuum isotropic exchange constant, $M_s = 1\times10^5$ \SI{}{\ampere\per\meter} is the saturation magnetisation, $D = 7.4\times10^{-5}$ \SI{}{\joule\per\meter\squared} is the isotropic DMI constant, $\mathbf{H_a}$ is the applied magnetic field which was set with a constant magnitude $\|\mathbf{H_a}\| = 4\times10^5$ \SI{}{\ampere\per\meter} applied mostly along the $z$-direction (out of plane), tilted by 1\degree~towards the $y$-direction to break symmetry. These values are the experimentally determined magnetic parameters of \ce{Cu2OSeO3}\cite{Zhang2018c}. To ensure we had an initial state within an energy minima, the state was relaxed using the \texttt{MinDriver()} method. After the state was relaxed, we used the \texttt{TimeDriver()} method, which relaxes the system taking into account the magnetisation dynamics which are governed by the Landau-Liftshitz-Gilbert (LLG) with spin-transfer torque from the Zhang-Li (ZL) model:

\begin{equation}
    \frac{\textnormal{d}\mathbf{m}}{\textnormal{d} t} = -\gamma\mathbf{m}\times\mathbf{H_{eff}} + \alpha(\mathbf{m}\times\frac{\textnormal{d}\mathbf{m}}{\textnormal{d}t})
    + \gamma\beta( \epsilon(\mathbf{m}\times\mathbf{m_P}\times\mathbf{m})-\epsilon'(\mathbf{m}\times\mathbf{m_P}),
\end{equation}

where $\gamma=2\times10^5$ m A$^{-1}$ s$^{-1}$ and $\alpha=0.5$ are the default values gyromagnetic ratio and damping parameters respectively. In a small number of cells within the barrier region, the gyromagnetic ratio was set to 0 to approximate the effects of pinning required for the localised state \cite{Leonov2017}. $\beta = 0.5 $ is the current density ($j_x$ = 100 A m$^{-2}$) dependant ZL damping parameter, $\epsilon, \epsilon'$ and $\mathbf{m_P}$ are the spin transfer terms and electron polarisation direction. Time-dependant micromagnetic simulations were done using a step-size of $2\times10^{-11}$ s, with every 100 steps saved for a total of 501 frames. Movies of the entire simulations can be found in the Supplementary Information: [URL will be inserted by the editor]

\section*{Data availability}
The data that support the findings of this study are available from the
corresponding authors upon reasonable request.

\section*{Acknowledgements}
This work was supported by the UK Skyrmion Project EPSRC Programme Grant (No.
EP/N032128/1). The SANS beamtime at the SINQ facility at the Paul Scherrer Institut
was awarded under proposal ID. 20211257. The magnetometry measurements were performed at the I10 support laboratory at Diamond Light Source and we are grateful for the technical assistance given by Mark Sussmuth. We thank the STFC’s ISIS Neutron and Muon Source for funding to cover PhD student travel expenses.

\section*{Author contributions}
S.H.M and P.D.H. conceived the project and designed the experiments. D.M. and G.B. produced the \ce{Cu2OSeO3} sample. S.H.M, M.T.L and J.S.W aligned and prepared the sample for the SANS experiment. S.H.M, M.T.L and D.A.V remotely performed and analysed the SANS measurements together with J.S.W onsite. S.H.M derived the free energy equations. S.H.M and M.T.L designed and performed the micromagnetic simulations. S.H.M. and M.T.L wrote the paper. All authors discussed the results and commented on the manuscript.

\section*{Competing interests}
The authors declare no competing interests.

\end{document}